\documentclass[preprint,showpacs,preprintnumbers,amsmath,amssymb]{revtex4}

\usepackage{textcomp}
\usepackage{makeidx}
\usepackage{amsmath}
\usepackage{subfigure}
\usepackage{amssymb}
\usepackage{hyperref}
\usepackage{graphicx}
\usepackage{epstopdf}
\usepackage{color}

\begin{document}

\title{Black hole at Lovelock gravity with anisotropic fluid.}

\author{Rodrigo Aros}
\email{raros@unab.cl}
\affiliation{Departamento de Ciencias F\'isicas, Universidad Andr\'es Bello, Av. Rep\'ublica 252, Santiago,Chile}
\author{Danilo Diaz}
\email{danilodiaz@unab.cl}
\affiliation{Departamento de Ciencias F\'isicas, Universidad Andr\'es Bello, Av. Rep\'ublica 252, Santiago,Chile}
\author{Alejandra Montecinos}
\email{alejandramontecinos@unab.cl}
\affiliation{Departamento de Ciencias F\'isicas, Universidad Andr\'es Bello, Av. Rep\'ublica 252, Santiago,Chile}
\author{Milko Estrada}
\email{mestrada@udla.cl}
\affiliation{Instituto de Matem\'atica, F\'isica y Estad\'istica, Universidad de las Am\'ericas, Manuel Montt 948, Providencia,Santiago, Chile}

\date{\today}

\begin{abstract}
In this work a new family of black hole solutions in Lovelock gravity is discussed. These solutions describe anisotropic fluids which extend to the spatial infinity. Though far from the horizon their geometries approach some previously known black holes solutions the location of the horizons differ. Furthemore, although the masses of these solutions match the masses of those previously known black holes, their temperatures and entropies differ.
\end{abstract}

\maketitle

\section{Introduction}

During the last decades several branches of physics have considered models in higher dimensions. In this scenario there have been an intense interest in the study alternative theories of gravity. Although the number of possible new theories of gravity is very large, if it is even finite, Lovelock gravities are particular as, despise containing higher power of the Riemann curvature, they still have second order equations of motion \cite{1}. In a matter of speaking, these theories can be casted, in more than one sense, as natural generalizations of the Einstein gravity for $d>4$, see for instance \cite{Zwiebach:1985uq,Banados:1993ur}.

The action principle of Lovelock gravities in $d$ dimensions is the sum, with arbitrary coefficients, of the lower dimensional topological invariants. For instance in $d=5$ this theory corresponds to the addition of Einstein Hilbert term -plus a cosmological constant- and the Gauss Bonnet term. As expected Lovelock gravities have for ground states states constant curvature manifolds, but in general there can be more than a single constant curvature manifold ground state. This in principle could generate unstable ground states in the space of solutions. Furthermore, it is expected that through dynamical evolution the geometry can jump between those different constant curvature manifolds.

Black hole solutions of the Lovelock gravities (in vacuum) have been extensively studied, for example \cite{Banados:1993ur,Crisostomo:2000bb,14,15} and reference therein. Unfortunately solutions in the presence of matter are by far rarer \cite{2,9,3}.

It is well known that the presence of matter fields can be cumbersome for gravity. In general, due to the non-linearities of gravity, any model that considers matter fields is difficult, if not impossible, to solve analytically for an arbitrary field configuration. Furthermore, usually in the literature to consider the presence of matter field, with few exceptions, is restricted to cosmological models and in general those considered are homogeneously and isotropically distributed. A classical exception of this is the static pressureless ball of dust which is a model for a static planet (or star). Although the exterior geometry of this model coincides with a Schwarzschild geometry this model can not describe a black hole as no horizon can arise from it. In \cite{7} the idea of an isotropic matter distribution around a particular point is considered to describe a non standard cosmology inside a black hole. This model gives rise to a (spherically) symmetric geometry around that point whose line element, in Schwarzschild coordinates, can be written by
 \begin{equation}
ds^2=- \Big ( 1- \frac{2m(r)}{r} \Big ) dt^2 + \frac{dr^2}{1- \dfrac{2 m(r)}{r}}+ r^2 (d\theta^2 + \sin^2 \theta d\phi^2 ). \label{5}
\end{equation}
where
\begin{equation}\label{4}
m(r) = 4 \pi \int_0^r \rho(r) r^2 dr = \frac{M}{2} \left( 1- e^{-\frac{r^3}{r_*^3}} \right),
\end{equation}
provided the energy density is given by $\rho(r) =  e^{- \frac{r^3}{r_*^3}}$ with $r^3_*=\alpha^2 M$ \cite{7}.

Since $\rho(r)$ extends to the radial infinity ($r=\infty$) it is not possible to define an exterior (Schwarzschild) metric in this case. However, it is direct to check that for $r \gg r_*$ this space tends to a Schwarzschild space. It is worth to notice as well that for $r \ll r_*$ the space behaves as a de Sitter space. This solution was latter generalized in \cite{17} by adding a cosmological constant term. In this case the line element is given by
\begin{align}
ds^2=- \left( 1 \pm \frac{r^2}{l^2}-\frac{2m(r)}{r}  \right) dt^2 + \frac{dr^2}{\left( 1 \pm \frac{r^2}{l^2}-\frac{2m(r)}{r}  \right)}
 + r^2 (d\theta^2 + \sin^2 \theta d\phi^2 ),\label{7}
\end{align}
where the cosmological constant is given by $\Lambda = \mp  \frac{6}{2 {l}^2}$. As expected for $r\gg r_*$ this space behaves as a Schwarzschild (Anti) de Sitter solution.

In this work we will generalize this result for Lovelock gravities with a single cosmological constant. First we will describe in some detail the Lovelock theories of gravity. Next, we will display solutions for a fixed Lovelock theory in the presence of that energy density. Finally we will study some features of the thermodynamics of these solutions.

\section{ Lovelock Gravity}

As mentioned above the action principle of Lovelock gravity in $d$ dimensions is the addition the lower dimensional topological invariants \cite{1,Banados:1993ur}. The Lagrangian can be written as
\begin{equation}
L \sqrt{-g} = \sqrt{-g} \sum_{p=0}^N \alpha_p L_p, \label{9}
\end{equation}
where $\alpha_p$ is a coupling constant. $N=\frac{d}{2}-1$ for even $d$ and $N=\frac{d-1}{2}$  for odd $d$. $L_p$ is defined by
\begin{equation}\label{Ln}
L_p = (d-2p)!\delta^{A_1 \ldots A_{2p}}_{C_1 \ldots C_{2p}} R^{C_{1} C_{2}}_{A_{1} A_{2}}\ldots R^{C_{2p-1} C_{2p}}_{A_{2p-1} A_{2p}},
\end{equation}
where $R^{C D}_{A B}$ is the Riemann tensor and
\[
\delta^{A_1\ldots A_n}_{C_1 \ldots  C_n} = \left|\begin{array}{ccc}                                                            \delta^{A_1}_{C_1}& \ldots & \delta^{A_1}_{C_n} \\
\ldots & \ldots & \ldots \\
\delta^{A_n}_{C_1} & \ldots & \delta^{A_n}_{C_n}
\end{array}
\right|
\]
is the generalized $n$-antisymmetric Kronecker delta. The first two terms in this series are proportional to
\begin{itemize}
\item  $L_1 \propto R $,
\item $L_2 \propto R^{ABCD}R_{ABCD}-4R^{ACBD}R_{ACBD}+ R^{AB}_{AB}R^{CD}_{CD}$.
\end{itemize}
It can be noticed that $L_2 = L_{GB}$ is the Gauss Bonnet density, where $R^{AB}=R^{ACB}_{\hspace{3ex}C}$ and $R=R^{AB}_{\hspace{2ex}AB}$ are the Ricci tensor and scalar respectively. In addition one can define $L _0 \propto 1$ giving rise to the cosmological constant provided $\alpha_0 \propto - 2 \Lambda$.

\section{Static and spherically symmetric solutions with anisotropic fluid}

As a first step only one of the terms of the Lovelock action principle will be considered to which a cosmological constant will be added. In this case the equations of motion are the generalization of the Einstein equations given by
\begin{equation}
 G_{AB}^{(n)} = T_{AB} \label{24}
\end{equation}
where
\begin{equation}\label{25}
G^{(n) AB} =  \frac{\delta }{\delta g_{AB}} \left((\alpha_n L_n + \alpha_0 L_0)\sqrt{g}\right)
\end{equation}
For simplicity in this work it will be considered a normalization such that
\[
\alpha_0 = \frac{(d-1)(d-2n)}{d} \frac{\alpha_n}{l^{2n}}.
\]
with $l^2>0$.

Our aim, as mentioned above, is to study spherical symmetric static solutions. For this it will be considered a metric in Schwarzschild coordinates defined by
\begin{equation}\label{22}
ds^2 =-f(r) dt^2+ \frac{dr^2}{f(r)} + r^2 d \Omega^2_{D-2}.
\end{equation}
The energy momentum tensor, given the symmetry, has the form $T^{A}_{\hspace{1ex}B} = \mbox{diag} (-\rho, p_r, p_\theta, p_\theta, ...)$. The strong energy condition determines that $\rho(r)>0$ and $\nabla_{A} T^{AB}=0$ yields
\begin{equation}
\frac{dp_r}{dr} +\frac{d-2}{r}(p_r-p_\theta ) =0, \label{18}
\end{equation}
which implies $p_\theta = \frac{r}{d-2} \frac{d}{dr}p_r + p_r$. Now, by considering the ansatz Eq.(\ref{22}) the relevant equations of motion of $G_{AB}^{(n)} = T_{AB}$ are
\begin{eqnarray}
 \rho(r)r^{d-2} &=& \alpha_n (d-2n)(d-1)!\left((d-1)\frac{r^{d-2}}{l^{2n}} - \frac{d}{dr} ( r^{d-2n-1}(1-f(r))^n )\right) \textrm{ and }\label{26} \\
 - p_r(r)r^{d-2} &=& \alpha_n (d-2n)(d-1)!\left((d-1)\frac{r^{d-2}}{l^{2n}} - \frac{d}{dr} ( r^{d-2n-1}(1-f(r))^n )\right)  \label{27}
\end{eqnarray}
From these equations it is direct to observe that $\rho=-p_r$. This establishes that this is an anisotropic fluid. It is worth to mention at this point that black holes solutions with the presence of anisotropic fluids have been studied in \cite{7,8,9,10,16,17}. The case $d-2n-1=0$ will be excluded to be discussed elsewhere. It can be noticed that Eq.(\ref{26}) can be readily integrated. Following \cite{7} it is natural to define
\begin{equation}\label{mass}
m(r) = \int_0^r \rho(r) r^{d-2} dr,
\end{equation}
which determines, upon integration, that
\begin{equation}\label{34}
\frac{l^{2n}}{\alpha_n(d-2n)(d-1)!} m(r)  =  r^{d-2n-1}(1-f(r))^n + \frac{r^{d-1}}{l^{2n}}.
\end{equation}

In order to invert this last relation, and to obtain $f(r)$, is necessary to consider the positivity of the argument and take only a particular branch of the $n$ branches of the $n$-root. For $n$ even the argument must always be positive in order to avoid imaginary values (and so to define a well posed metric). In practice the solution for $n$ even is only well defined for certain ranges of $r < \infty$. This implies that it is not possible to define an asymptotical AdS region in this case. This does not rule out the solution but restricts is interpretation. This will be discussed elsewhere.

On the other hand, for $n$ odd ($n=2k+1$) the situation is far simpler as it is always possible to choose the root of $(-1)^{1/{2k+1}}=-1$ and therefore \begin{equation}\label{f}
f(r) = 1+ \frac{r^2}{l^2}\left(1-\frac{l^{4k+2}}{\alpha_n(d-2n)(d-1)!}\frac{m(r) }{r^{d-1}}\right)^{\frac{1}{2k+1}}.
\end{equation}
It can be noticed that is well defined for $r\rightarrow \infty$. At this point usually $\alpha_n$ is chosen to simplify the expression, but as it will be shown below, this is unnecessary.

\section{Energy Density}

Now, in order to obtain an explicit solution is necessary to provide a mass density ($\rho(r)$). Ideally this density must be more realistic than the constant density pressureless ball of dust but simple enough to define an analytic mass function $m(r)$ (in Eq.(\ref{mass})). For reasons that will clear later on we choose the toy model
\begin{equation}\label{massDensity}
\rho(r) = \frac{d-2n-1}{d-2n}\frac{M}{V_{eff}}e^{- (r/R)^{d-1}},
\end{equation}
which is the generalization to $d$ dimensions of the one proposed in \cite{7} for $n=1$ and $d=4$. Here $R$ plays the role of an \emph{effective} radius. $V_{eff} = \frac{R^{d-1}}{d-1} \Omega_{d-2}$, where $\Omega_n$ is the area of $n$-sphere, corresponds to the effective volume of a ball of radius $R$ \footnote{This does not correspond to volume of a real ball of radius R defined from the geometry by a factor $e^{\lambda}$}.

First, it must be noticed that the strong energy condition determines that $M>0$. This mass density is defined such that
\begin{equation}\label{43}
m(r) =  \frac{(d-1)}{\alpha_n}\frac{d-2n-1}{d-2n}\frac{M}{\Omega_{d-2}} \left( 1- e^{- (r/R)^{d-1}}\right).
\end{equation}
For simplicity it convenient to define $\bar{M} = \frac{(d-1)}{\alpha_n}\frac{d-2n-1}{d-2n}\frac{M}{\Omega_{d-2}}$. Traditionally $\alpha_n$ can be chosen such that $\bar{M} = 2M$ but here this will be avoided.

Although it is direct to observe that this mass function rapidly converges into $\bar{M}$ as a function of $r$ its introduction is due to the fact that this density allows $f(r)$, under some particular conditions, to vanish. In turn, this defines the existence of black hole solutions unlike the usual constant density static pressureless ball of dust model.

\section{Limits}
In the next subsections we will discuss the different limit this solution presents. First, it must be emphasized that this solution is well defined for $0<r<\infty$ provided $n=2k+1$, thus we will restrict the analysis to this case. For instance, and unlike usual black holes, the curvature invariants of the geometry are well defined for that entire range and thus the geometry is smooth any value of $r$ in particular for $r=0$ and $r=\infty$. Therefore, this solution has a well defined asymptotically locally AdS region at $r\rightarrow \infty$.

To analyze the limit one must recognize the presence of three independent parameters in the solution; $M$, $R$ and $l$. In addition one must recognize that any two limits must commute in order to define a proper solution. For instance the mass must be insensible to any of limits in $l$ or $R$.

\subsection{ $r \ll R$ }

For $r \ll R$ is direct to notice that
\[
\lim_{r/R \rightarrow 0} \frac{m(r)}{r^{d-1}} = \frac{\bar{M}}{R^{d-1}}
\]
and thus the geometry tends to a constant curvature manifold with an effective cosmological constant proportional to 
\[
\frac{1}{l_{eff}^2} = \frac{1}{l^2}\left(1-\frac{\bar{M} l^{4k+2}}{R^{d-1}}\right)^{\frac{1}{2k+1}}.
\]
It is quite interesting that the effective cosmological constant may vanish, be positive or negative depending on the relation between $\bar{M}$ and $R$. Obviously this is totally different from any black hole solution where a singularity is expected at $r=0$. This feature shares with the static pressureless ball of dust model and is also similar to the result obtained by Dymnikova \cite{7} for $n=1$ and $d=4$.

It is very tempting, given the smoothness of the geometry together with presence of a potentially small but positive cosmological constant in the region $r/R \sim 0$, to try to connect this with a cosmological model for our universe. This is even more striking as this geometry is immersed in a negative cosmological constant environment.

\subsection{ $r\gg R$ }
In this case, for $r\gg R$, the effective mass of equation \eqref{43} is
\[
\lim_{r/R \rightarrow \infty} m(r) \approx \bar{M},
 \]
and thus the solution approximates
\begin{equation}\label{fprime}
\lim_{r/R \rightarrow \infty} f(r) = 1 + \frac{r^2}{l^2}\left(1-\frac{ \bar{M}l^{4k+2}}{r^{d-1}}\right)^{\frac{1}{2k+1}}.
\end{equation}
which corresponds to the family of black hole solutions obtained in \cite{14,15}.

As previously mentioned, it is not possible to call Eq.(\ref{fprime}) the exterior geometry as the mass density extends to $r\rightarrow \infty$. This merely demonstrates that the geometry for $r/R$ large enough becomes basically indistinguishable from the analogous black hole solution.

\section{Thermodynamics}

\subsection{Mass}

Following the standard methods one can define the mass of this family of solutions in terms of the Noether charge associated with the Killing vector $\partial_{t}$ in Eq.(\ref{22}). In principle one should consider as well to propose an action principle for the matter fields in order to define a second contribution to the expression of the Noether charge. However, also in general that second part of Noether charge usually does not contribute at the asymptotical region. In the case at hand, as the mass density is rapidly damped as $r$ increases, it will be assumed that the part of Noether charge associated with matter field action does not contribute to the asymptotical value of the Noether charge.

As mentioned in \cite{Aros:1999kt,Kofinas:2008ub} since the space is asymptotically locally AdS a regulator is necessary to compute the Noether charges. This regulator arises as a topological term added to the action principle. This is completely equivalent to holographic regularization. The introduction of this topological term regulates the action principle as well. The mass in this case, see appendix \ref{ComNC},
\begin{equation}\label{MassMilko}
  Q(\partial_t) = M + E_0
\end{equation}
where $M$ was defined in Eq.(\ref{43}), $E_{0} =0$ for even dimensions and the vacuum energy for odd dimensions \cite{Mora:2004rx}. The arise of a vacuum energy is feature of odd dimensional asymptotically locally AdS spaces. This confirms that the mass parameter corresponds to total mass of the solution as expected \footnote{This is the reason for the chosen normalization}.

\subsection{Temperature and Entropy}
As mentioned above, the function $f(r)$ may vanish for certain values of the parameters, which, from the form of the metric, defines the existence of a Killing horizon. For $n=2k+1$ (odd) $f(r)$ can be always positive, have two zeros and even has a double zero. In order to fix some ideas is useful to recall that $R$ and $l$ can be considered fixed parameter, as they determine the radius value where most of the mass/energy is located and the curvature of the asymptotical space.

The case with two zeros defines a space time bounded by the values of $r = r_{\pm}$. Unfortunately the numerical values of $r_+$ and $r_-$ are determined by the transcendental equation
\[
\frac{ l^{4k+2} \bar{M}\left(1 -e^{-(r_\pm/R)^{d-1}}\right)}{r_{\pm}^{d-3}} = 1 + \frac{r^{2}_\pm}{l^2}.
\]
or equivalently
\[
\bar{M} =\frac{r^{d-3}_{\pm}}{l^{4k+2}}\frac{\left(1 + \frac{r^{2}_\pm}{l^2}\right)}{\left(1 -e^{-(r_\pm/R)^{d-1}}\right)}
\]
It must be emphasized that the zeros of $f(r)$ differ from those defined by Eq.(\ref{fprime}). In fact Eq.(\ref{fprime}) defines a single horizon. This implies that, although both solutions share the same asymptotical behavior the solutions in Eq.(\ref{f}) represents a new and independent family of black holes solutions.

\begin{figure}[h]
  \centering
  \includegraphics[width=3in]{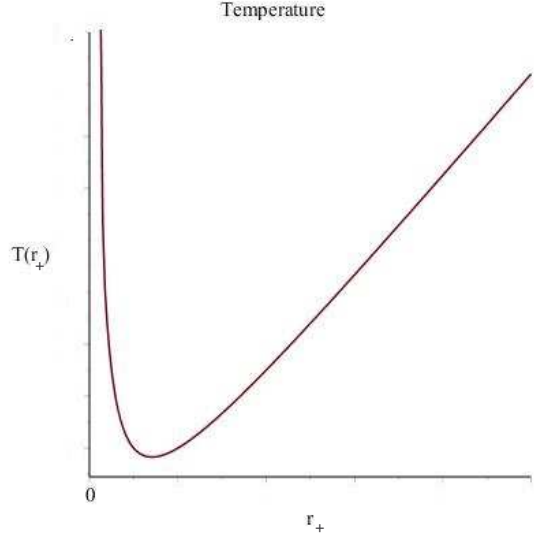}
  \caption{The temperature as a function of the $r_+$}\label{Temperature}
\end{figure}

The simplicity of the metric allows to determine some of thermodynamical properties. Considering $r=r_{+}$, the largest zero of $f(r)$, the temperature of the corresponding horizon is given by
\begin{eqnarray}
  T &=& \frac{1}{4\pi} \left. \frac{d}{dr} f(r) \right|_{r=r_{+}}\nonumber
 \\
   &=& \frac{1}{4\pi n} \left(\frac{d-2n-1}{r_+} + (d-1)\frac{r_+^{2n-1}}{l^{2n}} - \left(1+\frac{r_+^{2n}}{l^{2n}}\right)\frac{e^{-(r_+/R)^{d-1}}}{(1-e^{-(r_+/R)^{d-1}})}\frac{r_+^{d-2}}{R^{d-1}}\right) \label{TemperatureWeird}
\end{eqnarray}
It can shown as well that this temperature $T\rightarrow +\infty$ for either $r_+\rightarrow 0$ or $r_+\rightarrow \infty$, thus has in general a minimum, see for instance fig.\ref{Temperature}. Furthermore, $T$ can vanish as well (for $r_-=r_+$). Unfortunately since the relations are determined by transcendental equations it is not possible to define the temperature as a function of the mass of the solution. Furthermore, the relation may not be a function at all.

The form of the space time together with the action principle for gravity allow to define a canonical ensemble for the thermodynamics of these solutions,see for instance \cite{Aros:2001gz}. This, in turn, allows to compute the entropy as part of the Noether charge on the horizon following the original Wald's approach or its extensions. The entropy in this case is given by
\begin{equation}\label{Entropy}
  S = \frac{1}{T} \left. Q(\partial_t) \right|_{r=r_+}
\end{equation}
where $ Q(\partial_t)$ is the expression of the Noether charge associated with action principle defined by Eq.(\ref{Ln}) together with the regulator. Due to the topological pedigree of the regulator \cite{Kofinas:2008ub} arises a correction to Wald's expression by a local term evaluated at the horizon which does not depend on the parameters of the solution but $l$. The final result in $d$ dimensions for a $L_n$ is given by
\begin{equation}\label{EntropyFE}
  S = \alpha_n n (d-2)! r_{+}^{d-4k-2} \Omega_{d-2} + S_0
\end{equation}
where $S_0$ is a finite term independent of the value of $r_+$. It can noticed that this entropy does not follow an area law for a power $r^{4k}_+$ but defines an increasing function of $r_+$.

Unfortunately, the study of evolution of the evaporation requires of a numerical approach due to $r_+$ cannot be determined analytically. This is beyond the scope of this work and thus it will be analyzed elsewhere.

\section{Black hole Scan}

The solutions displayed above correspond to the cases of Lovelock gravities where the Lagrangian can be expressed as $L'_n =\alpha_n (R^n + \alpha' \Lambda)$. The presence of a single negative cosmological constant in these cases is manifest. However, there is an additional form to have a single cosmological constant. Roughly speaking this is obtained if the equation of motions have the form
\begin{equation}\label{BHS}
  \frac{\delta }{\delta g_{AB}} L\sqrt{g} \sim ((R + l^{-2})^n)^{AB}.
\end{equation}
This can be obtained provided the parameters $\alpha_p$ of Lovelock Lagrangian Eq.(\ref{9}) are given by \cite{Crisostomo:2000bb}
\[
\alpha^n_p = \left\{\begin{array}{cl}
                    \frac{\alpha_n}{d-2p}\binom{n}{p} &\textrm{ for } 0 \leq p \leq n\\
                    0 &\textrm{ for } n < p \leq N
                  \end{array}
\right.,
\]
where $\alpha_n$ is a global coupling constant. Their static (and vacuum) black hole solutions have been studied in \cite{Crisostomo:2000bb,Aros:2000ij}.

The generalization of these solutions to a density of mass of the form (\ref{massDensity}) is straightforward. In fact the solution can be obtained from the equation
\begin{equation}\label{BHSSolution}
  \alpha_n (d-2n)(d-1)! \frac{d}{dr}\left(r^{d-1}\left[\frac{\gamma - f(r)}{r^2} + \frac{1}{l^2}\right]^{1/n}\right) = \rho(r) r^{d-2}.
\end{equation}
Here it was introduced the generalization for a transverse geometry of constant curvature $\gamma$, see \cite{11,Aros:2000ij}, of the spherical transverse geometry in Eq.(\ref{22}) ($\gamma=1$). The solution is given by
\begin{equation}\label{BHSGsolution}
  f(r) = \gamma + \frac{r^2}{l^2} - \left(\frac{m(r)}{r^{d-2n-1}}\right)^{1/n}.
\end{equation}
where $m(r)$ is given by Eq.(\ref{43}).

As can be noticed from Eq.(\ref{BHSGsolution}) that in general there is no restriction for the dimension $d$ nor for the power $n$ in the Lagrangian. On the other hand, it is direct to notice that $f(r)$ may vanish for a certain value of parameters of the solution. This implies that for certain values of the parameter this solution corresponds to black holes. See below.

\subsection*{Limits of this solution}

This solution rapidly, as $(r/R) \rightarrow \infty$, recover the vacuum solution, \emph{i.e.},
\[
\lim_{(r/R)\rightarrow \infty} f(r) \simeq \gamma + \frac{r^2}{l^2} - \left(\frac{ \bar{M}}{r^{d-2n-1}}\right)^{1/n}
\]
as expected. It is worth to recall that the mass of the vacuum solutions for $d-1=2n$ and $\gamma=1$ is actually $\tilde{M} = M + 1/2$ . This is the case of the spherical symmetric Chern Simons black hole. The shift in the mass is due to the elimination of naked singularities from the spectrum of the (vacuum) solutions. In the case at hand this does not occur as no singularities at $r=0$ are presented. To observe that is enough to notice that the invariants of the geometry are finite at $r=0$. Furthermore,
\[
\lim_{r/R\rightarrow 0} f(r) = \gamma + r^2 \left(\frac{1}{l^2} - \left(\frac{\bar{M}}{R^{d-1}}\right)^{1/n}\right).
\]
which determines a smooth flat or (A)dS geometry at $r=0$ depending on the values of the parameters.

\subsection*{Mass and Thermodynamics}

It is straightforward to prove that the mass of these solutions can be obtained as the Noether charge associated with the Killing vector $\partial_t$. In general this requires a regularization \cite{Kofinas:2008ub}. The final result is given
\begin{equation}\label{MassBHS}
  Q(\partial_t)^{\infty} = M \Sigma_\gamma + E_0
\end{equation}
where $\Sigma_\gamma$ is the unitary area of the $d-2$ dimensional transverse section. $E_0 =0$ for even dimensions and corresponds to the vacuum energy for odd dimensions. This is in complete analogy with vacuum case but for Chern Simons case.

\begin{figure}[h]
  \centering
  \includegraphics[width=3in]{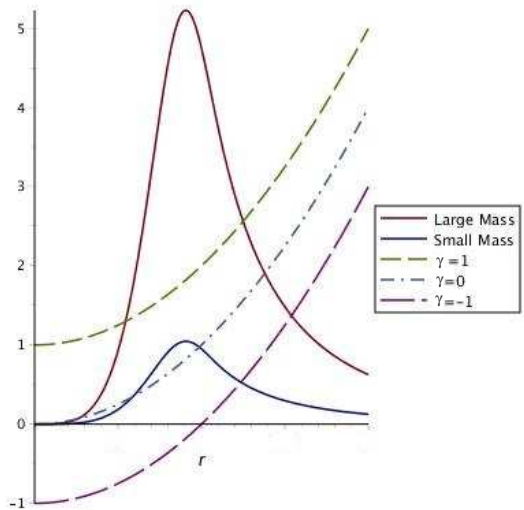}
  \caption{Plot of different values of $M$ and $\gamma$ to determine $f(r)=0$}\label{DiffMass}
\end{figure}

The differences with vacuum case, as previously, arise due the location of the horizons. In this case these are located at the values of $r$ which satisfy the transcendental equation  \begin{equation}\label{BHSzeros}
  \gamma + \frac{r^2}{l^2} = \left(\frac{\bar{M}\left(1 - e^{-(r/R)^{d-1}}\right)}{r^{d-2n-1}}\right)^{1/n}
\end{equation}
It is direct to prove that for $\gamma=1$ $f(r)$ have at most two zeros. For $\gamma=0$ there is always a solution at $r=0$ and up to three solutions. For $\gamma=-1$ there is always a single solution. This is depictured in Fig.(\ref{DiffMass}).

For the analysis of the thermodynamics it will be considered only the outer horizon whose location is defined by $r=r_+$. This determines that the mass parameter $\bar{M}$ can be obtained as
\begin{equation}\label{BHSMassParameter}
  \bar{M} = \frac{r_{+}^{d-2n-1}}{2\left(1 - e^{-(r_{+}/R)^{d-1}}\right)}\left(\gamma + \frac{r_{+}^2}{l^2}\right)^{n}
\end{equation}
The temperature of this horizon satisfies
\begin{equation}\label{BHStemperature}
4\pi T(r_+) = \frac{2r_+}{l^2} + \frac{1}{n}\left[\frac{(d-2n-1)}{r_+} - \frac{(d-1)r_+^{d-2}}{2 R^{d-1}} \frac{e^{-(r_+/R)^{d-1}}}{\left(1 - e^{-(r_{+}/R)^{d-1}}\right)}\right]\left(\gamma + \frac{r_+^2}{l^2}\right)
\end{equation}

In complete analogy with the previous case, the entropy can be computed by following the generalization of Wald's prescription. This yields
\begin{equation}\label{BHSEntropy}
  S = \alpha_n (d-2)! \left(\frac{r_+}{l}\right)^{d-2}\Sigma_\gamma \left[ \frac{n}{d-2} F\left([1-n,1-d/2],[2-d/2],-\frac{\gamma l^2 + r^2_+}{r^2_+}\right)\right] + S_0,
\end{equation}
where $\Sigma_\gamma$ is the area of the transverse section. It is matter of fact that Eq.(\ref{BHSEntropy}) is same expression obtained for the black hole solutions in \cite{Aros:2000ij}. The difference is contained in value of $r_+$ for a given value of $M$. $S_0$ is due to topological term added to regularize the action principle \cite{Kofinas:2008ub}, thus it is independent of the value of the mass, or in general of any parameter of the solution but $l$. It is also direct to check that for large $r_+$ this entropy always approach an area law.

It can be noticed that for $n=1$ the usual area law is recovered but this also occurs for $\gamma=0$ for any value of $n$ or $d$. It is direct to check that for $\gamma=1$ and $\gamma=0$ these functions are monotonically increasing functions of $r_+$ for any value of $n,d,l$. Far more interesting is the fact that $S_{\gamma=-1}(r_{+})$ can be negative which, as obviously cannot occur, defines should forbidden regions in the space of parameters. Unfortunately, as previously, due to $r_+$ cannot be determined analytically there are several aspect in addition to this last, as the study of evolution of the evaporation, which require of a numerical approach. Although interesting this will be considered for next work.

\section{Conclusions}

In this work some new static black hole solutions for some particular cases of Lovelock gravities have been found. These solutions, though correspond to non vacuum solution, for large $r$ reproduce the corresponding vacuum (black hole) solutions. In particular the mass of the solution reproduce the mass of the corresponding vacuum black hole solution. On the other hand, the solutions for $r\approxeq 0$ behave as maximally symmetric space whose radii of curvature depend on the value of exterior negative cosmological constant and the rest of the parameter of the solution.

One very interesting result of the ananlysis of this toy model is the fact that it has been established the possibility that two different families of black holes, who share the same asymptotical mass can have different thermodynamical properties. In this way the presence of matter fields outside of black hole can modify the location of horizon itself.

\acknowledgements

This work was partially funded by grants FONDECYT 1151107,1131075,1140296 and UNAB DI-735-15/R.

\appendix

\section{Computing the Noether charges}\label{ComNC}

In this particular case, after a straightforward computation, the regulated Noether charge is given by
\begin{equation}\label{NC}
\lim_{r\rightarrow \infty}  Q(\partial_t)  = \lim_{r\rightarrow \infty} \alpha_n (d-2)! \left( \int \frac{df(r)}{dr} (1 -f(r))^{n-1} r^{d-2n}d\Omega_{d-2} \right)
\end{equation}
For the sake of clarity it can be defined $f(r) = 1-g(r)^{\frac{1}{n}}$ which implies that
\begin{equation}\label{Mass}
\lim_{r\rightarrow \infty} Q(\partial_t) = \lim_{r\rightarrow \infty} \alpha_n (d-2)!\left(\frac{1}{n}\int \frac{dg(r)}{dr} r^{d-2} d\Omega_{d-2}  \right)
\end{equation}
In the case at hand
\[
g(r) = - \frac{r^2}{l^2} + \frac{n}{\alpha_n (d-2)!}\frac{M}{r^{d-2}}(1 - e^{-(r/R)^{d-1}})
\]
where $M$ was defined in Eq.(\ref{43}). The next step is to introduce the regulator, whose role is, roughly speaking, to eliminate the $r^{d-1}$-term presented in the expression above. The final result is
\[
\lim_{r\rightarrow \infty} \left. Q(\partial_t) \right|_{\textrm{Reg}} = M + E_0
\]
Here $E_0 =0$ for even dimensions and the vacuum energy for odd dimensions \cite{Mora:2004rx}.

\end{document}